\documentclass[aps,prl,10pt,superscriptaddress,twocolumn,floatfix]{revtex4-1}

\usepackage{natbib}
\usepackage[utf8]{inputenc}
\usepackage{xcolor}

\definecolor{linkblue}{RGB}{31,119,180}
\usepackage[
  unicode,
  colorlinks,
  citecolor=blue,
  linkcolor=linkblue,
  urlcolor=linkblue,
  bookmarks=true,
  bookmarksopen=true,
  bookmarksopenlevel=3,
  bookmarksnumbered=true]{hyperref}
\usepackage{graphicx}
\usepackage{amsmath}
\usepackage{amssymb}
\usepackage{lineno}
\usepackage{bm}
\usepackage{color}
\usepackage{graphicx}
\usepackage{amsmath}
\usepackage{braket}
\usepackage{changes}
\usepackage{booktabs}

\begin{document}

\title{The solution to the ``Einstein-Podolsky-Rosen paradox''}

\author{Roman~Schnabel}
\email{roman.schnabel@uni-hamburg.de}
\affiliation{Institut f\"ur Laserphysik \& Zentrum f\"ur Optische Quantentechnologien, Universit\"at Hamburg, Luruper Chaussee 149, 22761 Hamburg, Germany}

\date{\today}

\begin{abstract}
In 1935, Albert Einstein, Boris Podolsky, and Nathan Rosen (`EPR') reported on a thought experiment that they believed showed that quantum theory provided an incomplete description of reality. Today we know that quantum theory is a complete and correct description of Nature (in flat space-time). 
The EPR thought experiment is predicted by quantum theory and has since been experimentally confirmed.
However, EPR experiments have no physical pictorial explanation, and are considered part of a ``paradox''. 
Here I resolve the ``paradox''. I show independently of any interpretation of quantum theory that EPR (thought) experiments 
are a direct and compelling consequence of the existence of true randomness and the conservation of energy. 
It becomes obvious why EPR (thought) experiments allow for the precise prediction of truly random measurement values. Local hidden variables are not motivated.
\end{abstract}

\maketitle

\vspace{-2mm}
\section{Introduction}  \label{sec:1}
\vspace{-2mm}
When the quantum theory was formulated in the 1920s, one of its most prominent detractors was Albert Einstein. Together with his two co-authors Boris Podolsky and Nathan Rosen, he brought his criticism to the point in 1935, when they published a manuscript entitled \textquotedblleft \emph{Can Quantum-Mechanical Description of Physical Reality Be Considered Complete?}\textquotedblright \cite{Einstein1935}. It is today one of Einstein's most cited works. 
They tried to find out the answer to their question by first formulating a \textquotedblleft\emph{reasonable criterion}\textquotedblright\,for `physical reality':
\begin{quote}\vspace{-1mm}
\textquotedblleft \emph{If, without in any way disturbing a system, we can predict with certainty (i.e.,\;with probability equal to unity) the value of a physical quantity, then there exists an element of reality corresponding to that quantity.}\textquotedblright \hspace*{\fill} $[{\rm \!\;I\!\;}]$
\end{quote}\vspace{-1mm}
$[{\rm \;\!I\;\!}]$ is a proposal for a sufficient criterion of EPR's term `reality'. 
Fortunately, we do not need to discuss the term `reality', because the second column of the first page of Ref.\,\cite{Einstein1935} provides another sufficient criterion. If we combine both criteria, the term `reality' is bypassed and we get the actual criterion that EPR considered sufficient for proving incompleteness of a theory
\begin{quote}\vspace{-1mm}
\textquotedblleft \emph{If, without in any way disturbing a system, we can predict with certainty (...) the value of a physical quantity --- but this value has no counterpart in physical theory --- then the theory is incomplete.}\textquotedblright \hspace*{\fill} $[\!\;{\rm II}\!\;]$
\end{quote}\vspace{-1mm}
EPR continued their publication with the famous EPR thought experiment, in which values of \emph{uncertain} physical quantities could be predicted with \emph{certainty}. They concluded: 
\textquotedblleft \emph{While we have thus shown that the wave function does not provide a complete description of the physical reality, we left open the question of whether or not such a description exists. We believe, however, that such a theory is possible.}\textquotedblright ~
Niels Bohr and Erwin Schr{\"o}dinger immediately published counter statements \cite{Bohr1935,Schroedinger1935}. 
They did not reject the thought experiment, but expressed their conviction that quantum theory was complete. 

Apart from this, EPR's work did not attract further attention for more than 25 years. 
It changed in the 1960s when John Bell \cite{Bell1964} succeeded in formulating an inequality that made it possible to test experimentally whether it is possible to complete a theory so that it contains counterparts for all local measurement values that can be precisely predicted in EPR experiments. The apparently missing counterparts are today called `local hidden variables' \cite{Clauser1969}. 
Since the 1970s and up to recent years, so-called `Bell tests' have repeatedly drawn a great deal of attention    \cite{Freedman1972}\nocite{Aspect1981,Maddox1991,Weihs1998,Ansmann2009,Gisin2009,Giustina2013,Hensen2015,Shalm2015}-\cite{Moreau2019}. Using a subclass of the EPR-entangled states -- the so-called Bell-entangled states --  they (i) proved the possibility of making predictions within the ranges of minimum uncertainty products, but at the same time (ii) refuted the possibility of adding local hidden variables creating `local realistic theories'. Today, they are said to have refuted `local realism' \cite{Aspect1981,Weihs1998,Groblacher2007}. 
The combination of (i) and (ii) is called `quantum nonlocality' \cite{Ghosh1986}, which so far is not understood \cite{Maddox1991,Gisin2009}.

For many physicists, the possibility of precise predictions without local hidden variables was surprising, and the question arose whether the Bell tests had have loopholes. 
But to promote potential loopholes, strange and implausible assumptions about Nature have to be made. Anyway, all three main potential loopholes have since been closed by experiments \cite{Weihs1998,Giustina2013,Hensen2015,Shalm2015}. 

EPR did believe in the existence of local hidden variables and did not use the word `paradox'. It was introduced by Schr\"odinger \cite{Schroedinger1935} and appeared again in the 1960's, when John Bell also formulated his inequality \cite{Bell1964}. 
The meaning of the term ``EPR paradox'' has changed over time and is only vaguely defined.
The EPR paradox is certainly not about a violation of Heisenberg's uncertainty principle, because on one subsystem only one quantity is ever measured in EPR experiments \cite{Einstein1935}.
It is also not about information that propagates faster than light, see also \cite{Deutsch2000}.
A clear description from today's perspective is required. %
Given the results of the Bell tests, the ``EPR paradox'' is the contradiction of two (seemingly) correct facts.
\begin{quote}\vspace{-1mm}
Fact 1: The EPR paper starts from a reasonable criterion $[{\rm \!\;I\!\;}]$ (more precisely:\,$[{\rm \!\;II\!\;}]$) and builds on it an argumentation in the framework of the correct EPR thought experiment and comes to the compelling conclusion that 
quantum theory needs to be completed (by local hidden variables), at least for all EPR entangled states.\\[2mm]
Fact 2: The Bell tests prove that there are no local hidden variables, 
at least for a subclass of the EPR-entangled states -- the so-called Bell-entangled states.
\end{quote}\vspace{-1mm}
The EPR criterion, in the form of $[{\rm \!\;I\!\;}]$ or $[{\rm \!\;II\!\;}]$, is the only weak point in the EPR work and thus the starting point to solve the paradox. But it has been completely unclear \emph{why} the EPR criterion should be wrong. 
I note that disproving the existence local hidden variables, (Fact 2) does not solve the EPR paradox. First of all, it is rather part of it. Second, the proven absence of local hidden variables together with the proven possibility of making predictions lead to `quantum nonlocality', whose physics is not understood \cite{Maddox1991,Gisin2009}.
\\[-2mm] 

Here I first present a simple thought experiment that exposes EPR's sufficient criterion for the incompleteness of a theory as false. I then consider the experimental explicit realisations of the EPR thought experiment and provide a quantum physical explanation of why we can predict the measured value of a physical quantity with certainty (i.e.~with probability equal to one) although the measured value is truly random due to quantum uncertainty and local hidden variables do not exist.
My explanation is exclusively based on the existence of true randomness and quantized interaction in combination with classical correlations and boundary conditions such as the conservation of energy. 
My explanation exposes also Einstein's 1927 argument for the incompleteness of quantum theory incorrect \cite{Harrigan2010}.\\  [-3mm]

\vspace{-2mm}
\section{True randomness: ``no reality''} \label{sec:2} %
\vspace{-2mm}
In almost all situations is (quantum) physics not able to predict the exact amount of energy that is redistributed when two physical systems interact, nor the exact time of a spontaneous decay. Even if identical measurements are performed on identical systems, there are nevertheless different measurement results. 
The statistics fulfill all the properties of true random statistics. The widths of the probability distributions are the known quantum uncertainties.
One could assume that the respective systems, however, carry `local hidden variables' that predetermine exactly what happens when in each individual case. With this assumption, measurement statistics could still look truly random, although they are not. 
But this assumption would be wrong, since the experimental violations of Bell inequalities \cite{Freedman1972,Aspect1981,Weihs1998,Giustina2013,Hensen2015,Shalm2015} have excluded the existence of local hidden variables.
They prove that there can be no reason why a measurement gives a predetermined exact value within quantum uncertainty.
They thus prove that quantum uncertainties describe ranges of truly random measurement values.
Within that range, a local measurement quantity is said to have no pre-existing local reality.\\
An instructive example is given by measuring the exact times of radioactive decays of individual atoms.
The recorded times are randomly and uniformly distributed before or after the half-life, regardless of when the actual monitoring begins.
Since local hidden variables are excluded, the times of radioactive decays are \emph{truly} random within the exponentially decaying probability curve. 
To clarify the meaning of the term "true randomness", let us make the following statement: Measured values that are truly random occur for no reason.
Note that `true randomness' has been also called `intrinsic randomness' \cite{Bera2017}.

I suspect that the majority of physicists agrees on the previous paragraph.
However, there is a large group of physicists who go beyond physics and explore on a metaphysical level whether there are still ways to avoid true randomness as well as the absence of reality. This is about `interpretations' of quantum physics. 
My manuscript is \emph{not} about interpretations.\\ [-3mm]

\vspace{-2mm}
\section{The EPR thought experiment} \label{sec:3}
\vspace{-2mm}
By 1935, EPR were convinced that quantum theory itself provided an argument against the existence of true randomness. The EPR thought experiment is a statistical experiment and it obeys quantum theory. 
It consists of many repetitions of measurements of two orthogonal, non-commuting observables on identical systems to show that quantum theory allows individual measurement results to be inferred (`predicted') with a precision better than the minimum Heisenberg uncertainty product. If the EPR criterion quoted on the first page were correct, the EPR thought experiment would disprove quantum uncertainty as the realm of true randomness (and expose quantum theory as incomplete).
In colloquial terms, EPR argued that \textquotedblleft \emph{Something that can be inferred cannot have happened truly randomly.\textquotedblright}
In their \emph{thought} experiment, the inference was arbitrarily precise, and EPR concluded that there was no true randomness at all in Nature. 
EPR proposed that quantum theory must be supplemented by local hidden variables that replace true randomness with randomness that is only apparent to us. 
\begin{figure}[t!!!!!!!!!!!!!!!!]
 \center 
     \vspace{0mm}
    \includegraphics[angle=0,width=4cm]{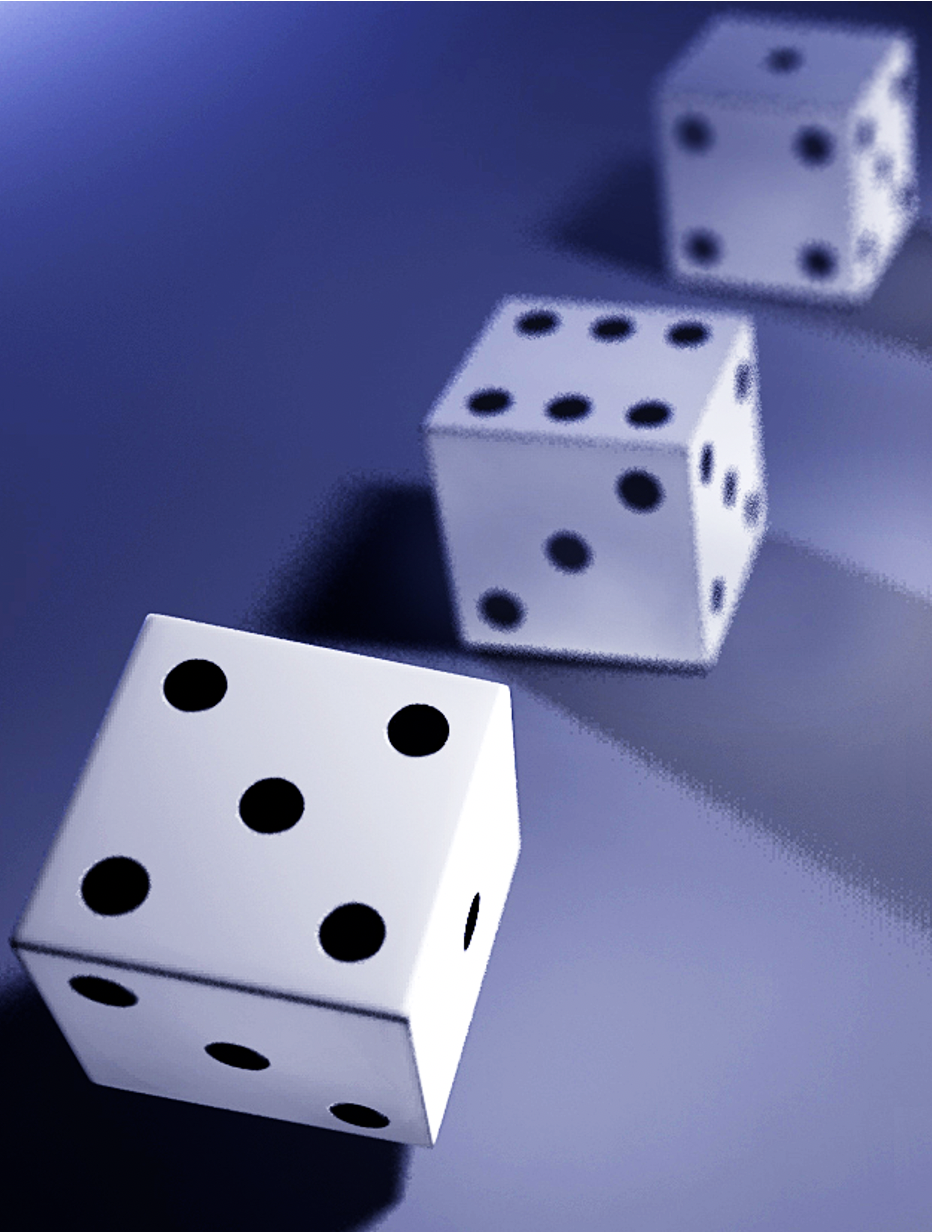}
    \vspace{-1mm}
    \caption{{\bf Inferring measurement outcomes} -- Let us assume that many identical dice come to rest in a mutually independent truly random fashion. The latter makes the dice at least \emph{semiclassical} if not even \emph{`quantum'}, because there is no true randomness in classical physics. 
    The observation shown shall correspond to an ensemble measurement of the physical quantity `number at the top'. It results in a truly random series `A'. Another truly random series `B' results from the `number at the bottom'.  
    Without looking at the undersides, we know that the numbers are 6, 1 and 2, read from back to front. 
    The illustrated semiclassical thought experiment strictly refutes the seemingly logical criterion $[{\rm \!\;II\!\;}]$ . 
    First, the theory of truly random dice cannot and is not expected to predict any number, but is nevertheless complete. Second, it is obvious that the two correlated truly random results allow for an inference without interaction. The fact that a pair of values (A$_i$, B$_i$) is produced by \emph{one} object, indeed corresponds to the situation in quantum physics, where two entangled subsystems are \emph{inseparable}, i.e.~form a single object. 
      What makes my thought experiment different from earlier \cite{Bell1981} is the assumption of true randomness.    
      Credit: Figure by Alexander Franzen. 
}
    \label{fig:1}
\end{figure}

Demonstrating the EPR thought experiment, the precision of the inference (`prediction') does not need to be perfect. 
It only needs to be more precise than the mini\-mal range of true randomness according to the quantum theory. If a quantum uncertainty has a Gaussian shape, the minimal value for its standard deviation depends directly on the standard deviation of the orthogonal quantity,  
as given by the minimum of Heisenberg uncertainty principle \cite{Heisenberg1927,Kennard1927,Weyl1927,Robertson1929}.
The latter's most cited form considers the position $x$ and the momentum $p$ of the centre of mass of any physical system and reads
\begin{equation}
\Delta \hat x \cdot \Delta \hat p \geq \hbar/2 \, ,
\label{eq:1}
\end{equation}
where $\Delta \hat x$ and $\Delta \hat p$ are the respective quantum uncertainties in terms of standard deviations, 
and $\hbar$ is the reduced Planck constant. 
It is important to realise that the only role of the Heisenberg uncertainty principle in the EPR thought experiment is to quantify a range of
guaranteed true randomness according to the quantum theory. For a given $\Delta \hat p$, the scatter of position measurement values of at least $\hbar/(2 \Delta \hat p)$ must be free of any determinism and thus exhibit true randomness. 
In contrast, measurement spectra with a product larger than the minimum in Eq.\,(\ref{eq:1}) may have a partially deterministic contribution. 
Anyway, if $\Delta p$ is not quantified, any value $\Delta x'$ could be generated by a deterministic mixture of perfect position-squeezed states. In this case, the measured $x$-values were not truly random at all.

The EPR thought experiment deals with pairs of subsystems `A' and `B' that are in a fully symmetric maximally position-momentum entangled state \cite{Schroedinger1935}.
The measurements of $\hat x_{\rm A}$, $\hat x_{\rm B}$, $\hat p_{\rm A}$, and $\hat p_{\rm B}$ clearly obey Eq.\,(\ref{eq:1}) with subscripts either `A' or `B'.
But whenever a pair is used for two position measurements, the difference of the two outcomes does not show any uncertainty. And whenever the two momenta are measured, the sum of the two outcomes does not show any uncertainty, i.e.
\begin{equation}
\Delta (\hat x_{{\rm A}} - \hat x_{{\rm B}})  =  \Delta (\hat p_{{\rm A}} + \hat p_{{\rm B}}) = 0 \, .
\label{eq:2}
\end{equation} 
Note that for the more general \emph{asymmetric} maximally position-momentum entangled state, either two positive or two negative factors should be added to either `A's' or `B's' quantities \cite{Reid1989}.
EPR correctly pointed out that a measurement of $\hat x_{{\rm A}}$ allows to precisely infer $\hat x_{{\rm B}}$, despite their non-zero uncertainties. 
The alternative measurement of $\hat p_{{\rm A}}$ allows to precisely infer $\hat p_{{\rm B}}$.
Starting from their criterion, this proves the existence of a local hidden reality that nullifies the true randomness of quantum theory, and they proposed to supplement quantum theory with local hidden variables that describe the reality they thought they had found.

\vspace{-2mm}
\section{The flaw in EPR's logic} \label{sec:4}
\vspace{-2mm}
The EPR work \cite{Einstein1935} is based on their seemingly logic criterion quoted on the first page here. 
My simple thought experiment in Fig.\,\ref{fig:1} makes clear that their criterion is false. Fig.\,\ref{fig:1} illustrates many identical {`quantum'} dice coming to rest in a \emph{truly} random fashion. The statistics of the observed numbers on the top face show the expected flat probability distribution of $1/6$ for the possible outcomes from one through six. A real experiment would at best yield pseudo-random numbers, because in reality, the individual results are mainly due to specific microscopic initial conditions and differences in the trajectories of the tosses. \\  
Visible in Fig.\,\ref{fig:1} are example results of the ensemble measurement at `A'. The ensemble measurement at `B' corresponds to the numbers on the bottom. Every two numbers of a pair become reality at the same time. We, as the observer of `A' cannot observe the numbers at `B', however, we can predict every single number with certainty. The reason is the strict constraint that opposite numbers add up to seven. Importantly, the numbers at `B' are just as random as the ones at `A'. 
I have just shown the following:
\emph{If, without in any way disturbing a system, we can predict with certainty (i.e.,\;with probability equal to unity) the value of a physical quantity, it is nevertheless possible that this value has no counterpart even in a complete theory. The reason is that the respective value has shown up as part of two perfectly correlated random processes.} 

Interestingly, the rigorous assumption of a true randomness leads to the fact that the actually classical cube becomes not only semiclassical, but even completely quantum mechanical (with two entangled numbers on opposite sides). The strict assumption of true randomness requires that the rolling, following which the cube comes to rest, is already understood as part of the measurement process. When the cube is \emph{rolling with respect to the environment}, as partially visible in (Fig.\,\ref{fig:1}), it already has sides coupled to the environment, albeit with a dynamics. The environment is a thermal bath, thus in a thermal state. This is not at the minimum of Heisenberg's uncertainty principle and therefore its influence on the cube is not truly random, but partially deterministic. True randomness requires a cube that has no sides that are somehow related to the environment. The only property of the quantum mechanical cube is the following: If it happens that it has opposite sides, then there are two numbers with the sum 7.

\begin{figure}[h!!!!!!!!!!!!!!!!]
 \center 
     \vspace{0mm}
    \includegraphics[width=6.2cm]{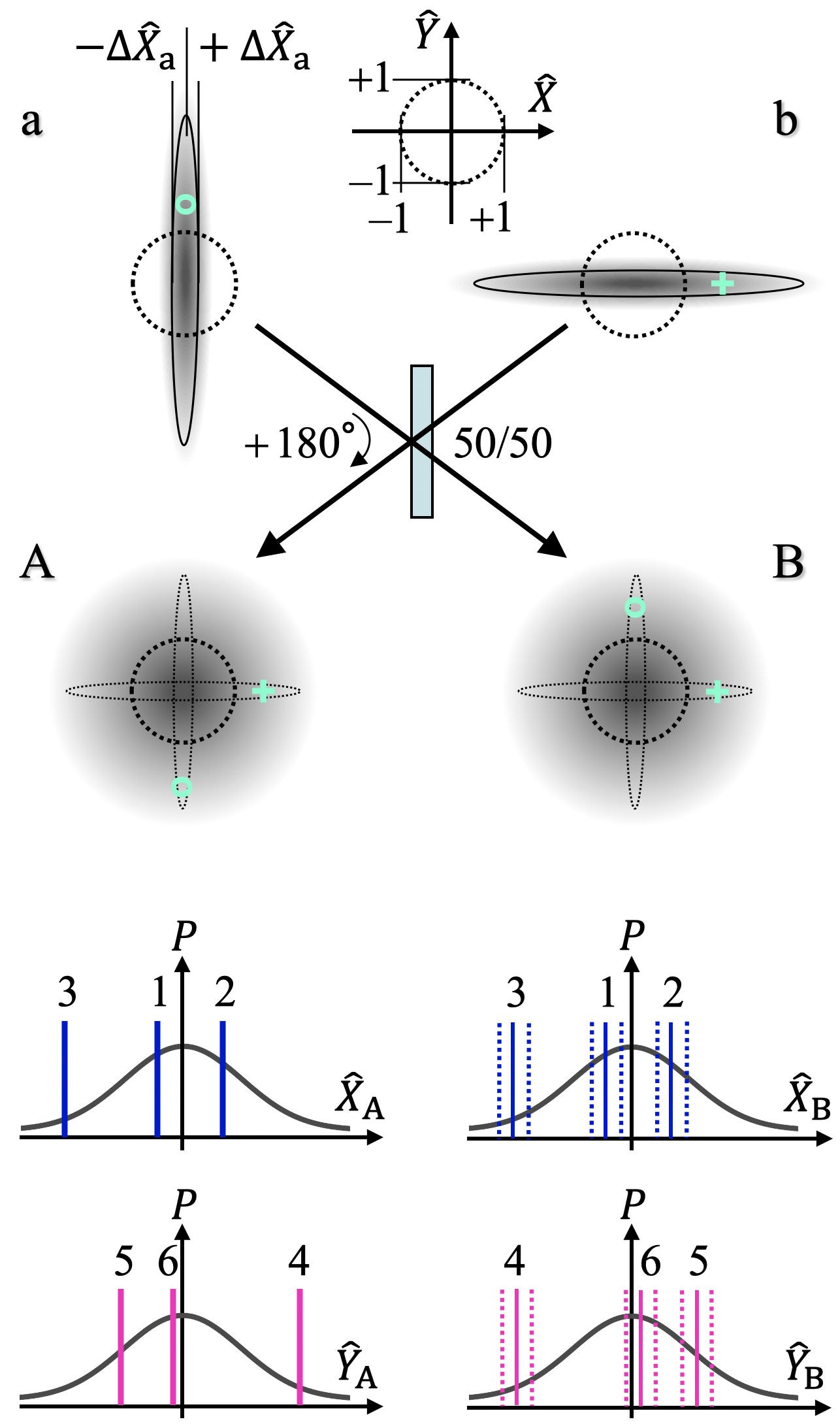}
    \vspace{0mm}
    \caption{{\bf Illustration of the completed EPR thought experiment} -- The upper half shows the emergence of (Gaussian) EPR entangled states A and B. Two independent input systems a and b in pure squeezed vacuum states (ellipses) mutually distribute their $\hat X_{\rm a,b}$-$\hat Y_{\rm a,b}\,$-phase space uncertainties in the course of superposition and balanced energy exchange represented by a 50/50 beam splitter. ${\bm \circ}$ and {\footnotesize \bf +} label the positive ranges of the input uncertainties. Dashed circles represent ground state uncertainties. The shapes of the entangled uncertainties A and B are due to the superposition principle. Their sizes and the $180^\circ$-phase space flip are enforced by energy conservation. The lower half of the figure illustrates ensemble measurements of $\hat X_{\rm A,B}$ and $\hat Y_{\rm A,B}$ on the output systems A and B revealing local Gaussian probability distributions. Since local phase space uncertainties represent ranges without local reality (local hidden variables do not exist), every single measurement result must be a truly random number. However, if the same observables are measured on A and B (all six example measurements from \#1 to \#6), every single measurement value at A allows a precise prediction of the measurement outcome at B, and vice versa. The remaining uncertainty of the prediction is given by the squeeze strength of the input uncertainties and indicated by the dashed lines in the measurement histograms of B. The figure clarifies why the apparently logical EPR criterion that a truly random event cannot be predicted is a fallacy. It thus solves the EPR paradox. Furthermore, the figure presents the physics of EPR entanglement and `nonlocal referenced realism' (see Sec.\,\ref{sec:6}) and explains why these must inevitably arise from the deterministic redistribution of quantum uncertainties, see main text.
}
    \label{fig:2}
\end{figure}

A feasible quantum-physical experiment that conveys the same insight as Fig.\,\ref{fig:1} is the well-known effect of spontaneous decay such as spontaneous pair production and radioactive decay. The spontaneously decay products appear truly random, as the word `spontaneous' suggests. 
But it is quite obvious that when one part is `born', one can immediately predict with certainty the existence of another part.
It becomes clear that there are situations in which truly random events or measurement results can be predicted (better: ``inferred'') with certainty. 
Another quantum-physical experiment is described in Fig.\,\ref{fig:2}. It finally solves the EPR paradox. \\[-3mm]

\vspace{-2mm}
\section{The solution to the EPR paradox in actual demonstrations} \label{sec:5}
\vspace{-2mm}
EPR did not realise that their thought experiment, including the possibility of making precise predictions about uncertain quantities, could be explained in a logical and pictorial way by reverting exclusively to quantum theory as it was established as early as 1935, i.e. without finding any apparent incompleteness of it.
Here, I provide this explanation. For this, I consider actually performed experiments that demonstrated the EPR paradox with quantum states having Gaussian uncertainties. The first EPR experiment with Gaussian uncertainties was performed with optical fields in 1992 \cite{Ou1992}. The type of EPR entangled states produced were subsequently used to demonstrate quantum teleportation \cite{Furusawa1998,Bowen2003}, to advance quantum imaging, magnetometry and interferometry \cite{Marino2009,Wasilewski2010,SteinlechnerS2013}, 
to implement one-sided device independent quantum key distribution \cite{Gehring2015}, and to research the fundamentals of the Heisenberg uncertainty principle \cite{Zander2021}.
In all quoted experiments, the EPR entangled quantities were the depths of amplitude and phase quadrature modulations of quasi-monochromatic laser light, the so-called quadrature amplitudes $\hat X_{f, \Delta f}$ and $\hat Y_{f, \Delta f}$ \cite{Schnabel2017}, where $f$ is the frequency of the modulation and $\Delta f$ its resolution bandwidth. An illustration of the quadrature amplitudes for monochromatic modulations ($\Delta f \rightarrow 0)$ can be found in the supplementary information of Ref.~\cite{Zander2021}. 
The quadrature amplitudes are proportional to optical field strengths but normalised to be dimensionless and in such a way that the corresponding Heisenberg uncertainty relation reads\\[-2mm]
\begin{equation}
\Delta \hat X \cdot \Delta \hat Y \geq 1 \, ,
\label{eq:3}
\end{equation}
where the subscript $f,\Delta f$ is now omitted.
In the above quoted experiments, the quantum correlations in the measurement values of the quadrature amplitudes of beam A and beam B were of finite strength but with canonical symmetry: $\hat X_{{\rm A}} \approx + \hat X_{{\rm B}} $ and $\hat Y_{{\rm A}}\approx - \hat Y_{{\rm B}} $ (or with swapped plus and minus signs).
With this symmetry, the EPR paradox and thus EPR entanglement is certified if the following inequality is violated \cite{Reid1988,Reid1989,Bowen2003a,Bowen2004}
\begin{equation}
\Delta (\hat X_{{\rm A}} - \hat X_{{\rm B}})  \cdot  \Delta (\hat Y_{{\rm A}} + \hat Y_{{\rm B}}) \geq 1 \, .
\label{eq:4}
\end{equation}
The stronger the violation, the more significant is the EPR paradox. 
The experiment reported in \cite{Eberle2013} reached a value below $0.18$ for the left-hand side of inequality (\ref{eq:4}).
The Heisenberg uncertainty relation in inequality (\ref{eq:3}) is not violated because inequality (\ref{eq:4}) contains conditional variances.

Fig. 2 shows how Gaussian EPR entanglement is produced to realise the EPR thought experiment. %
Required are two laser beams of identical wavelengths (a and b) that carry amplitude squeezed states  ($\Delta \hat X_{{\rm a}} \!<\! 1$) and phase quadrature squeezed states ($\Delta \hat Y_{{\rm b}} \!<\! 1$), respectively. Pure squeezed states \cite{Stoler1970,Walls1983,Breitenbach1997,Schnabel2017} have a pair of $\hat X$ and $\hat Y$ quadratures with Gaussian distributions whose standard deviations are at the lower limit of the Heisenberg uncertainty principle according to inequality (\ref{eq:3}). 
Notably, measurement values of $X$ and $Y$ of any two (pure or mixed) single-mode squeezed states do not violate inequality (\ref{eq:4}).
The two squeezed beams (a) and (b) are spatially overlapped on a beam splitter. The strength of the entanglement is maximized for a balanced splitting ratio.
The two beam splitter output beams (A) and (B) jointly carry an ensemble of an Gaussian EPR entangled pair.
(Gaussian EPR entangled states are also called `two-mode squeezed states'.) 
Notably, measurement values of $\hat X_{{\rm A}}$ and $\hat X_{{\rm B}}$ do violate inequality (\ref{eq:4}).

The beam splitter input-output relation for expectation values of optical fields has to obey energy conservation as well as the symmetry in the setup.
The input-output relation cannot change for a superposition of optical fields and are thus the same for quantum uncertainties. 
The two squeezed Gaussian input uncertainties are thus transferred to two Gaussian output uncertainties. 
The new standard deviations are weighted sums of those of the inputs, yielding for a balanced beam splitter $\Delta \hat X_{{\rm A}} = \Delta \hat X_{{\rm B}} = (\Delta \hat X_{{\rm a}} + \Delta \hat X_{{\rm b}})/\sqrt{2}$ and $\Delta \hat Y_{{\rm A}} = \Delta \hat Y_{{\rm B}} = (\Delta \hat Y_{{\rm a}} + \Delta \hat Y_{{\rm b}})/\sqrt{2}$. The quantum uncertainties of the resulting states `A' and `B' are illustrated in Fig.\,\ref{fig:2} by the two large blurred circles. %

Conservation of energy enforces a \emph{deterministic} way of transforming input uncertainties into output uncertainties. The sum of the absolute squares of the two input field amplitudes must be identical to the sum of the absolute squares of the two output field amplitudes (in the absence of loss channels). The energy is conserved if, in addition, a $180^\circ$ phase rotation applies to one of the reflected quadratures. %

The logical consequence is correlations in the two output beam uncertainties that are "narrower" than the phase space areas of the ground states of the individual beams `A' and `B'. %
(The emergence of correlations in the quantum uncertainties of `A' and `B' are pointed out in Fig.\,\ref{fig:2} by ${\bm \circ}$ and {\footnotesize \bf +}.)
Such correlations within quantum uncertainties are called `quantum correlations'. In Fig.\,\ref{fig:2}, the quantum correlations are equivalent to EPR entanglement and enable rather precise predictions that qualitatively correspond to that in the original EPR thought experiment.
The four Gaussian probability distributions at the bottom of the figure represent the local measurement spectra including the possibility to predict measurement outcomes at `B' via measurement results at `A'. For perfect input squeezing ($\Delta \hat X_{{\rm a}}, \Delta \hat Y_{{\rm b}} \rightarrow 0$) the prediction becomes arbitrarily precise. I note that the entire spectra of measurement results are due to quantum uncertainties, and thus truly random, although the uncertainty areas of the EPR entangled states are much larger than the lower bound of inequality (\ref{eq:3}). 
For perfect input squeezing, the  uncertainty areas of `A' and `B' approach infinity.

Fig.\,\ref{fig:2} completes the EPR thought experiment by adding the generation of the entangled EPR states.
The figure is not a simplified model of the EPR (thought) experiment, but a complete representation of the relevant quantum physics, 
because the phase space uncertainties in Fig.\,\ref{fig:2} are a complete description of the states and their evolution.

Fig.\,\ref{fig:2} solves the EPR paradox because it shows that the EPR criterion ($[{\rm \;\!I\;\!}]$) or ($[{\rm \;\!II\;\!}]$), respectively, can only be false. Assuming that quantum uncertainties represent regions in which measured quantities have no reality, i.e. measured values are truly random, the superposition of two squeezed uncertainties inevitably leads to the fact that locally truly random measurement values can be inferred with certainty. In other words, the said superposition inevitably leads to EPR entanglement. 

At the 1927 Solvay Conference, Einstein had presented a simpler thought experiment which he also believed showed that quantum theory was incomplete. An illustration can be found in Fig.\,8 of Ref.~\cite{Harrigan2010}: a mode in a one-electron Fock state is diffracted in two directions and the two parts are measured with perfect quantum efficiency. 
Quantum theory cannot predict at which detector the electron will be measured before a measurement is made, but the result at one detector (no electron or one electron) allows an accurate prediction of the result at the other detector (one electron or no electron). 
With the explanations of the EPR paper worked out here, it becomes immediately clear that also this thought experiment gives no reason to doubt the completeness of the quantum theory. After diffraction, the mode of the electron has a transversal spatial uncertainty which overlaps 50\% each with the surfaces of two measuring apparatuses. The localization of the electron within the uncertainty is truly random, so it is impossible for any theory to make a prediction in advance of a measurement. Since the interaction is quantized, it follows from conservation of energy that one measuring apparatus will measure exactly one electron if the other measures exactly none. The measurement result of Einstein's 1927 thought experiment is therefore the only possible one, if one considers the Heisenberg uncertainty principle as fundamental and thus measurement results within the uncertainties as truly random.

\vspace{-0mm}
\section{Summary and conclusions}  \label{sec:7}
\vspace{-2mm}
The experimental violations of Bell-type inequalities already proved that the 1935 work of Einstein, Podolsky, and Rosen was flawed. 
Here I make obvious why it was the basic assumption of EPR, given by the criterion quoted on the first page, that was flawed. 
I explain that EPR's `reality of a physical quantity' (or `realism') is synonymous with `no true randomness of measured values'. 
I propose a simple thought experiment with semi-classical dice (Fig.\,\ref{fig:1}) that makes it clear that the reality of a physical quantity does not necessarily follow from the possibility of predicting measurement results with certainty without disturbing the system. 
Explicitly, my thought experiment thus makes it clear that \emph{truly} random measurement values can nevertheless be predicted with certainty. 
I note that the EPR thought experiment is \emph{not} about a prediction concerning a point in time before a measurement, i.e.~before any part of the EPR entangled state has interacted with the environment. At the time of the `prediction', the dice have already come to rest. I suggest using the word `inference' instead of `prediction'.

EPR's condition for `reality' is not sufficient, i.e.~the following implication is \emph{false}

\vspace{1.5mm}\hspace{3mm} 
\emph{predictability $\Rightarrow$ reality\,$/$no true randomness}. 
\vspace{1.5mm}\\
In the presence of a complete theory, only the following implications are correct:

\vspace{1mm}
\hspace{3mm} \emph{predictability $\Leftarrow$ reality\,$/$no true randomness},
\vspace{1mm}

\hspace{3mm} \emph{unpredictability $\Leftrightarrow$ no reality\,$/$true randomness}.
\vspace{1.5mm}

Since the predictability of measured values is not a sufficient condition for local realism, the EPR thought experiment provides no motivation for local realism (equivalent to `local hidden variables' or `non-existence of true randomness'), as already refuted by the experimental violations of the Bell inequalities. %

My Fig.\,\ref{fig:2} represents a complete quantum mechanical illustration of the EPR thought experiment, including the physics of Gaussian EPR entanglement.  
The figure clarifies the following:
Assuming that quantum uncertainties represent regions in which 
measured values are truly random, the superposition of two squeezed uncertainties leads to two truly random measurement series with mutual correlations or anti-correlations, which represents the EPR gedanken experiment and thus EPR entanglement. 
I argue that EPR entanglement 
is in fact \emph{unavoidable} because the superposition of field amplitudes must be deterministic in order not to violate conservation of energy.  
My result complements the experimental violations of Bell inequalities, which already proved that quantum uncertainties represent regions in which measured quantities have no reality, i.e.~measured values are truly random.

I would like to draw the following conclusions from my work. %
Since my work solves the EPR paradox and reveals the physics of EPR entanglement, %
it also increases the probability of finding physical explanations for all remaining strange quantum phenomena. %

\vspace{8mm}
\begin{acknowledgments}
\textbf{Acknowledgments} --
RS thanks Mikhail Korobko and Ralf Riedinger for useful comments on the manuscript. This work was performed within the European Research Council (ERC) Project \emph{MassQ} (Grant No.~339897) and within Germany's Excellence Strategy -- EXC 2056 `Advanced Imaging of Matter', project ID 390715994 and EXC 2121 `Quantum Universe', project ID 390833306, which are financed by the Deutsche Forschungsgemeinschaft.
\end{acknowledgments}

\end{document}